\documentstyle[NATO,epsfig]{crckapb} 
\newcommand{\gcc}{{\rm g} \, {\rm cm}^{-3}}
\newcommand{\dmdt}{{\mbox{{\rm M}$_{\odot}$}} {\rm yr}$^{-1}$}

\newcommand{\mdot}{\mbox{$\dot{M}$}}

\newcommand{\lsim}{\raisebox{-0.3ex}{\mbox{$\stackrel{<}{_\sim} \,$}}}
\newcommand{\gsim}{\raisebox{-0.3ex}{\mbox{$\stackrel{>}{_\sim} \,$}}}
\begin{opening}
\title{MAGNETIC FIELDS OF NEUTRON STARS}
\author{Sushan Konar}
\institute{Inter-University Centre for Astronomy \& Astrophysics, Pune}
\author{Dipankar Bhattacharya}
\institute{Raman Research Institute, Bangalore}
\end{opening}
\begin{document}

\begin{abstract}
The evolution of the magnetic field is investigated for isolated as well as binary neutron stars. The overall nature of 
the field evolution is seen to be similar for an initial crustal field and an expelled flux. The major uncertainties of 
the present models of field evolution and the directions in which further investigation are required are also discussed 
in detail.
\end{abstract}

\section{Introduction}
There is no consensus regarding the generation of the magnetic field in neutron stars. The field could either be a fossil 
remnant from the progenitor star in the form of Abrikosov fluxoids of the core proton super-conductor (Baym, Pethick \& 
Pines 1969, Ruderman 1972, Bhattacharya \& Srinivasan 1995). Or it could be generated after the formation of the neutron star 
in which case the currents would be entirely confined to the solid crust (Blandford, Applegate \& Hernquist 1983). Evidently, 
the nature of the evolution would depend very much on the internal field configuration. Observations and statistical analyses 
of existing pulsar data, nevertheless, indicate that significant decay of magnetic field is achieved only if the neutron star 
is a member of an interacting binary (Bailes 1989, Bhattacharya 1991, Hartman et al. 1997). \\

\noindent The processes that are responsible for the field evolution in neutron stars in binaries are - a) expulsion of the
magnetic flux from the super-conducting core during the phase of propeller spin-down, b) screening of the field by accreted 
matter and c) rapid ohmic decay of the crustal field in an accretion-heated crust (for a review see Bhattacharya 2000). 
Diamagnetic screening of the field by accreted matter does not seem likely to have any long-term effect (Konar 1997) and we 
shall exclude it from the present discussion. The other models invoke ohmic decay of the current loops for a permanent decrease 
in the field strength. In either case, the effect of accretion is two-fold. The heating reduces the electrical conductivity and 
consequently the ohmic decay time-scale inducing a faster decay. At the same time the material movement, caused by the deposition 
of matter on top of the crust, pushes the original current carrying layers into deeper and denser regions where the higher 
conductivity slows the decay down. The mass of the crust of a neutron star changes very little with a change in the total mass; 
accretion therefore implies assimilation of the original crust into the super-conducting core. When the original current carrying 
regions undergo such assimilation, further decay is stopped altogether. Both the purely crustal model as well as the model assuming 
an expelled flux have been investigated by many authors (see Bhattacharya 2000). The important difference between our work and that 
of the other investigators lies in our assumption of a {\it flux freezing} upon the assimilation of the original current carrying 
layers into the super-conducting core.

\begin{figure}
\begin{center}{\mbox{\epsfig{file=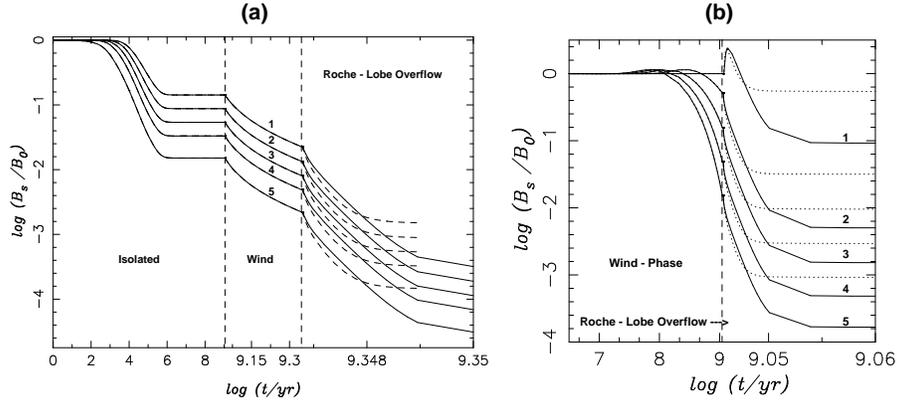,width=140pt,angle=-90}}}\end{center}
\caption{Evolution of the surface magnetic field in LMXBs. {\bf a)} For initial crustal currents. With an wind accretion rate of 
$\mdot = 10^{-14}$~\dmdt. Curves 1 to 5 correspond to initial current configurations centred at $\rho = 10^{13}, 10^{12.5}, 
10^{12}, 10^{11.5}, 10^{11} \gcc$. All curves correspond to an impurity content,$Q$ = 0.0. A standard cooling has been assumed for 
the isolated phase. The wind and Roche-contact phases are plotted in expanded scales. The dashed and the solid curves correspond 
to accretion rates of $\mdot = 10^{-9}, 10^{-10}$~\dmdt in the Roche-contact phase. {\bf b)} For an expelled flux. The dotted and 
the solid curves correspond to accretion rates of $\mdot = 10^{-9}, 10^{-10}$~\dmdt in the Roche contact phase. The curves 1 to 5 
correspond to $Q$ = 0.0, 0.01, 0.02, 0.03 and 0.04 respectively. All curves correspond to a wind accretion rate of $\mdot = 10^{-16}$~\dmdt.}
\label{f_lmxb}
\end{figure}

\begin{tabular}{|l|l|r|} \hline
\multicolumn{3}{|c|}{ TABLE I - {\bf Purely Crustal Field}} \\ \hline
system & final field and period & comment \\ \cline{1-3}
isolated & high field, long period & no significant field decay \\
radio pulsars & & in $10^9$~ years  \\ \cline{1-3}
HMXB & high field, long period & high-mass binary pulsars \\
& & and solitary counterparts  \\ \cline{2-3}
& low field, long period & not active as pulsars  \\ \cline{1-3}
LMXB & high field, long period & high field low-mass \\
& & binary pulsars \\
& & and solitary counterparts  \\ \cline{2-3}
& low field, short period & low field low-mass \\
& & binary pulsars \\
& & and solitary counterparts, \\
& & millisecond pulsars  \\ \cline{1-3}
\end{tabular}

\section{Nature of Field Evolution}
\subsection{Generic Features}
The qualitative features of field evolution, as outlined below, are similar for a) an initial crustal field and 
b) an expelled flux. 

\noindent {\bf Pure Ohmic Decay in Isolated Neutron Stars} (Konar 1997) \\
\noindent 1. A slow/fast cooling of neutron star implies a fast/slow decay; hence a low/high final field. \\
\noindent 2. Initial crustal currents concentrated at lower/higher densities gives rise to low/high final surface fields. \\
\noindent 3. Large impurity content makes the decay rapid and gives rise to smaller final fields. 

\noindent {\bf Accretion-Induced Field Decay in Accreting Neutron Stars} \\
\noindent (Konar \& Bhattacharya 1997 - KBI) \\
\noindent 1. In an accreting neutron star the field undergoes an initial rapid decay, followed by slow down and an eventual 
{\em freezing}.  \\
\noindent 2. A positive correlation between the rate of accretion and the final field strength is observed, giving rise to
higher final saturation field strengths for higher rates of accretion. 

\noindent {\bf Magnetic Field and Spin Period} -- We have investigated the nature of the final `magnetic field-spin period' 
combination. Our results agree well with the observations and are summarised in table-I (KBI, Konar \& Bhattacharya 1999a - KBII). 
The nature of field evolution is similar for the model of spin-down induced flux expulsion (Konar \& Bhattacharya 1999b - KBIII). 
Though there is one major difference as can be seen from fig.(\ref{f_lmxb}). To produce millisecond pulsars in LMXBs for an
expelled flux large values of impurities, in the prior-to-accretion original crust, are required. But this would result in 
extremely small surface fields in old isolated pulsars. This is in complete contrast to a purely crustal model and expectation 
from statistical analyses of pulsar data.  

\subsection{Ranges of Physical Parameters} 
\begin{figure}
\begin{center}{\mbox{\epsfig{file=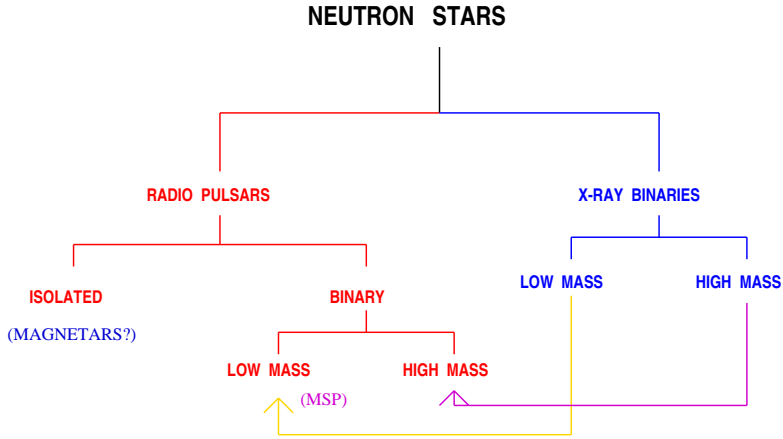,width=200pt,angle=-90}}}\end{center}
\caption{The arrows indicate the expected evolutionary link between X-ray binaries and binary radio pulsars.}
\label{f_paradigm}
\end{figure}

The paradigm of field evolution that have emerged out of observations, statistical analyses and theoretical expectations 
have been summarized in fig. (\ref{f_paradigm}) where the connection between the radio pulsars and their binary counterparts,
namely the X-ray binaries is indicated. In table-II we indicate the constraints on various physical parameters in the field 
evolution models placed by the requirement to match observed properties in a variety of systems (KBII, KBIII). The parameters 
discussed here are - the density at which the initial crustal current distribution is located ($\rho_c$), the impurity strength 
in the crust ($Q$), the duration of wind-accretion phase in different binary systems and the rate of accretion in the Roche-contact 
phase for LMXBs.

\begin{tabular}{|l|l|l|r|r|} \hline
\multicolumn{5}{|c|}{TABLE II - {\bf Constraints on Physical Parameters}} \\ \hline
parameter & model & system & requirement & parameter range \\ \cline{1-5}
$\rho_0$ & crustal & HMXB & high field & high $\rho_0$ \\ \cline{1-5}
$Q$ & crustal & isolated & no decay over& $Q \lsim 0.01$ for \\
& & radio & active pulsar & standard cooling, \\
& & pulsar & life-time & $Q \lsim 0.05$ for \\
& & & & accelerated cooling \\ \cline{2-5}
& expelled & LMXB & millisecond & $Q \gsim 0.05$ with \\
& flux & & pulsar & wind accretion, \\
& & & generation & $Q >> 1$ without \\
& & & & wind accretion \\ \cline{1-5}
duration of & crustal & HMXB & high field & short \\
wind accretion & & & & \\ \cline{1-5}
\mdot~in & crustal & LMXB & high field & Eddington rate \\
Roche-phase & & & & \\ \cline{1-5}
\end{tabular} 

\section{Uncertainties and Future Directions}
The results and conclusions stated above suffer from a number of uncertainties regarding the micro-physics of the neutron star.
as have been listed below. Moreover, a lot of the new theoretical results as well as the observational facts have recently become
available. In this section we mention some of the more important aspects that need to be incorporated in any future work on the 
evolution of the magnetic fields of neutron stars.

\noindent {\bf Thermal Behaviour} \\
\noindent 1. Isolated Phase - The present data can be made to fit scenarios with both a {\em slow} or an {\em accelerated} 
cooling. Therefore, it is not clear which is the correct cooling behaviour of an isolated neutron star. \\
\noindent 2. Accreting Phase - The crustal temperature corresponding to a given rate of accretion has not been determined with 
any degree of certainty. Also, the existing results are limited in their scope. \\
\noindent 3. Post-Accretion Phase - No calculation exists for the thermal behaviour of this phase at all.

\noindent {\bf Transport Properties} - Several factors affect the transport properties and hence both thermal and magnetic 
field evolution. Prominent among them is the change in the {\bf chemical composition} due to a) accretion and b) spin-down. 
Recently it has been shown that the impurity content of the accreted crust for near-Eddington accretion rates could be extremely
large (Schatz et al. 1999). This, along with a temperature inversion near or beyond the neutron drip (Brown 1999) might 
modify the transport properties significantly. Moreover the presence of dislocations, defects, non-spherical nuclei have so 
far not been taken into account in the calculation of transport properties. These are also expected to have an impact on the 
field evolution in isolated as well as binary pulsars.\\

\noindent {\bf Multi-polar Structure} - All of our and similar investigations have been based on an assumption of a pure dipolar 
model for the magnetic field. Though calculations for isolated neutron stars does not show any appreciable change in multi-polar 
structures (Mitra, Konar \& Bhattacharya 1999) - the situation would change in presence of accretion or a very strong magnetic field 
(Geppert et al. 1999) due to the importance of the Hall term requiring further investigation. \\

\noindent {\bf The Magnetar Question} - Amongst some of the more recent developments the magnetars pose a great challenge for the 
existing theories of field evolution since they require a very rapid field evolution in isolated neutron stars. Though some work have 
already been done in this area (Heyl \& Kulkarni 1998, Geppert et al. 1999) - more detailed investigation is needed.

\end{document}